\documentclass[twocolumn,prl,aps,showpacs]{revtex4}
\usepackage{epsfig}
\begin{document}  
\title{
Fringe Structure in the Phase-Space Dynamics of Atomic Stabilization in An
Intense Field }
\author{Jie Liu $\,^{1,2}$,Shi-gang Chen $\,^2 $,
 Baowen Li $\,^1$ and  Bambi Hu $\,^{1,3}$}
\affiliation{Department of Physics and Centre for Nonlinear Studies,
 Hong Kong Baptist University, Hong Kong}
\affiliation{Institute of Applied Physics and
Computational Mathematics, P.O.Box.8009,  100088 Beijing, China}
\affiliation{Department of Physics, University of Houston, Houston TX
77204, USA}

\begin{abstract}
An analytical expression of a Floquet operator, which describes the evolution
of a wave packet in combined atomic and an intense laser field,
 is obtained approximately in the stabilization regime.
  Both the classical and quantum versions
 of the Floquet operator are used to study the phase-space
dynamics of atomic stabilization, and   the   
 'fringe structure'  in the phase-space
 is clearly demonstrated. Furthermore, we investigate
 the  dynamical mechanism  and characters of this striking structure, and
 find that this structure is very closely related to the classical
 invariant manifolds. 
\end{abstract}
\maketitle

\section{INTRODUCTION}
Atomic stabilization means that  the
 ionization rate of an atom decreases with increasing  intensity of intense
 , high-frequency laser field, which is one of 
the most interesting nonperturbative
phenomena observed in laser-atom interaction and attracts much attention 
recently.
 This effect has been predicted by both quantum
 theories $[1]$ and classical   theories $[2]$, and has recently been
 verified experimentally$[3,4]$.
 In particular, Sundaram and Jensen [5] connect successively
 this exciting phenomenon with the chaotic dynamics. 

 More recently, by numerically solving time-dependent Schrodinger
 equation for one-dimensional model,
 some researchers have investigated the phase-space dynamics
  of intense-field stabilization and found the existence of
  a striking structure - fringe structure $[6,7]$.
These findings enrich greatly  the discussions on atomic stabilization and
 helpful to better understand the stabilization mechanism.
However, the  dynamical mechanism of the fringe structure 
is far from fully understood. 

 In this paper, using the approximate analytic expression of the evolution
 operator of the atom-laser system,
   we shall demonstrate the fringe
   structure from classical distribution and quantal Wigner distribution.
    Furthermore, in light of   nonlinear dynamics 
 we  shall discuss the dynamical mechanism  underlying this 
striking structure.
    As we shall see later that,  
    the  fringe
   structure associates closely  with  the ionization process in the
   regime of stabilization.

The  
one-dimensional model is widely used in studying the atom  
stabilization phenomenon $[1,2]$ in the strong-field laser-atom physics, 
its Hamiltonian is 
\begin{equation}
H(x,v,t)=\frac{v^2}{2}
+V(x)
-xF\cos\omega t \,\,\, ,
V(x)=-1/\sqrt{1+x^2} \,\,\, ,
\end{equation}
where $F, \omega $ are the  strength and frequency of
the oscillating electric field, respectively. In all calculations of this paper, 
the atomic units 
($m_e = e = \hbar =1 a.u.$) is adopted 
for simplicity.

Usually, the stabilization phenomenon is  discussed in the
Kramers-Henneberger (KH) frame, i.e.  the laser-atom
interaction is viewed from the stand  point of a free classical
electron oscillating in a lasert field .
By this way, the  Hamiltonian becomes
\begin{equation}
H_{KH}(q,p,t)= \frac {p^2}{2} + V(q-\alpha_0\cos\omega t) \,\,\, ,
\end{equation}
where $\alpha_0=F/\omega^2$ is the amplitude of an electron's quiver motion.
Notice that the wave function in the KH frame is related to that in
the length gauge by an unitary transformation.

This is a periodically driven system. With the help of a time-order exponential,
the formal evolution operator at $t>0$ can
be written as,
\begin{equation}
\hat{U}(t) = P [exp(-i\int_0^tdt'\hat{H}_{KH}(t'))] \,\,\, ,
\end{equation}
where $P$ is the time order operator.

The time evolution operator referring to one period $T$ is the
 so-called Floquet operator
 \begin{equation}
\hat{F}=\hat{U}(T) \,\,\, .
 \end{equation}
If the frequency of  the laser field is assumed
to be asymptotically high, we can  discard all but the
the time average of the KH potential, namely,
\begin{equation}
V_0(q)=
 \frac 1 T \int_0^T V(q-\alpha_0\cos\omega t) dt \,\,\, .
\end{equation}
Then  the  Hamiltonian becomes  time independent
and takes the form of 
\begin{equation}
\bar H_{KH} = \frac {p^2}{2} + V_0(q) \,\,\, .
\end{equation}
The eigenstates of this system is  called KH states.
So one is led  to the prediction that all KH states will
be stable.
However, in practical case of finite field frequency rather than
an asymptotically high frequency, Sundaram and Jensen pointed out that
the high order Fourier terms are also important and can not be neglected
for simplicity. 
 With the assumption that every coefficients of the Fourier expansion
 of $V(q-\alpha_0\cos\omega t)$ are equally important, the
 potential can be described by a periodic train of delta kicks as follows,
 \begin{equation}
 H_{KH}=\frac{p^2}{2}+V_0(q)\sum_{n=-\infty}^{+\infty}\delta(\frac t T - n) \,\,\, .
 \end{equation}
 In $[5]$ the validity of  the above assumption has been tested for a wide
 range of parameters.
 It is found that in the regime of atomic stabilization  the approximation
 can   provide a
 good qualitative description.
 Provided that the zero of time is chosen halfway between two
 consecutive kicks, in terms of (3) and (7), the Floquet operator is 
 \begin{equation}
 \hat{F} = \hat{R} \circ \hat{K} \circ \hat{R} \,\,\, ,
  \end{equation}
  where
  \begin{equation}
  \hat{R}=exp(-ip^2T/4) \,\,\, ,
  \end{equation}
  represents the half-periodic time evolution of a free  electron, and 
  \begin{equation}
  \hat{K}=exp(-iV_0T) \,\,\, .
  \end{equation}
  represents  an impulse  on momentum.

  The above evolution operator  
  describes the interaction between field and atoms .
  It is easy to show that the above  operator is
  time-reversal invariant.

\section{Classical Description of
Ionization Process}
The classical version of the above quantum Floquet operator (eq.8) takes form,
\begin{equation}
M=R\circ K\circ R \,\,\, ,
\end{equation}
where $R: (q,p) \rightarrow (q+pT/2,p)$,
propagates the trajectory for a half-period ahead along
the constant momentum;
 $K: (q,p) \rightarrow (q,p-T\frac{\partial V_0}{\partial q})$, describes
an impulse on the electron.
Obviously,   $M$ is area-preserving in phase plane.

By iterating  $M$ for a variety of initial conditions and
plotting the trajectories in the phase plane $(q, p)$, we can construct
a Poincare surface of section which provides an overview of the classical
dynamics.
The phase plots of $M$ demonstrate many different structures for 
different  parameters $F$ and $\omega$. 
A detailed analysis show that, for the case of stabilization the phase
structures can be classified into three main cases as we show in Figs. 1,2 
and 3.
Case one, as shown
in Fig.1,
the $P_1(0,0)$ is the unique 1-periodic elliptic
fixed point. Outside the regular island around
$P_1$, the phase plane
 is filled with chaotic trajectories. In case two (fig.2),
with changing the parameters
, this elliptic point $P_1$ undergoes a pitch  bifurcation to an
unstable hyperbolic fixed point and gives birth to
a pair of 1-periodic elliptic fixed points ($P_2, P_3$) locate at
$(\pm\alpha_0, 0)$ approximately. Outside  these two regular islands
the phase plane is full of stochastic orbits.
 Case three (fig.3), the $P_2$ and $P_3$ becomes unstable,
 and
  the phase plane is mainly filled by chaotic trajectories
 and  very small   fraction of regular islands.
 In term of classical description of ionization, the
 electrons initiated in the islands will never be ionized. The regular
 islands around high-periodic fixed points are too small compared
 with that of 1-periodic fixed points and contribute little to the
 confinement of the electrons. Therefore, the above analyses provide
 three typical parameters $(F, \omega)$ which are used  in following
 discussions on ionization suppression.

\begin{figure}[!htb]
\begin{center}
\resizebox *{8cm}{8cm}{\includegraphics*{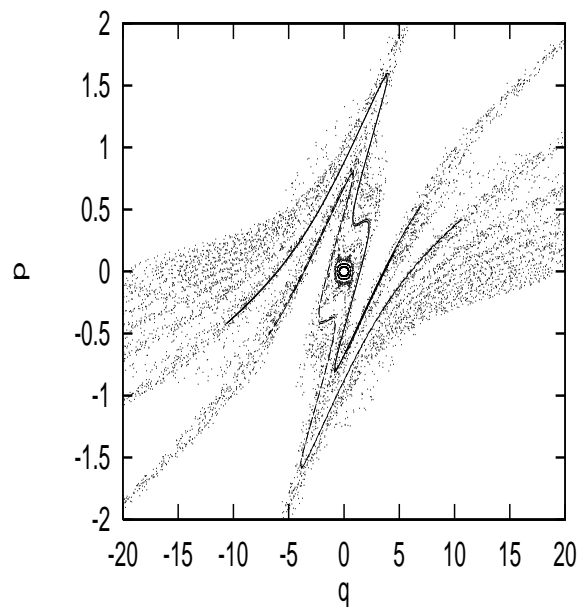}}
\end{center}
\caption{
Poincare surface of section for $ F=1.28$ and $\omega=0.8$ (Case 1).
Solid line gives the unstable manifolds.
}
\label{fig:fig1}
\end{figure}

\begin{figure}[!htb]
\begin{center}
\resizebox *{8cm}{8cm}{\includegraphics*{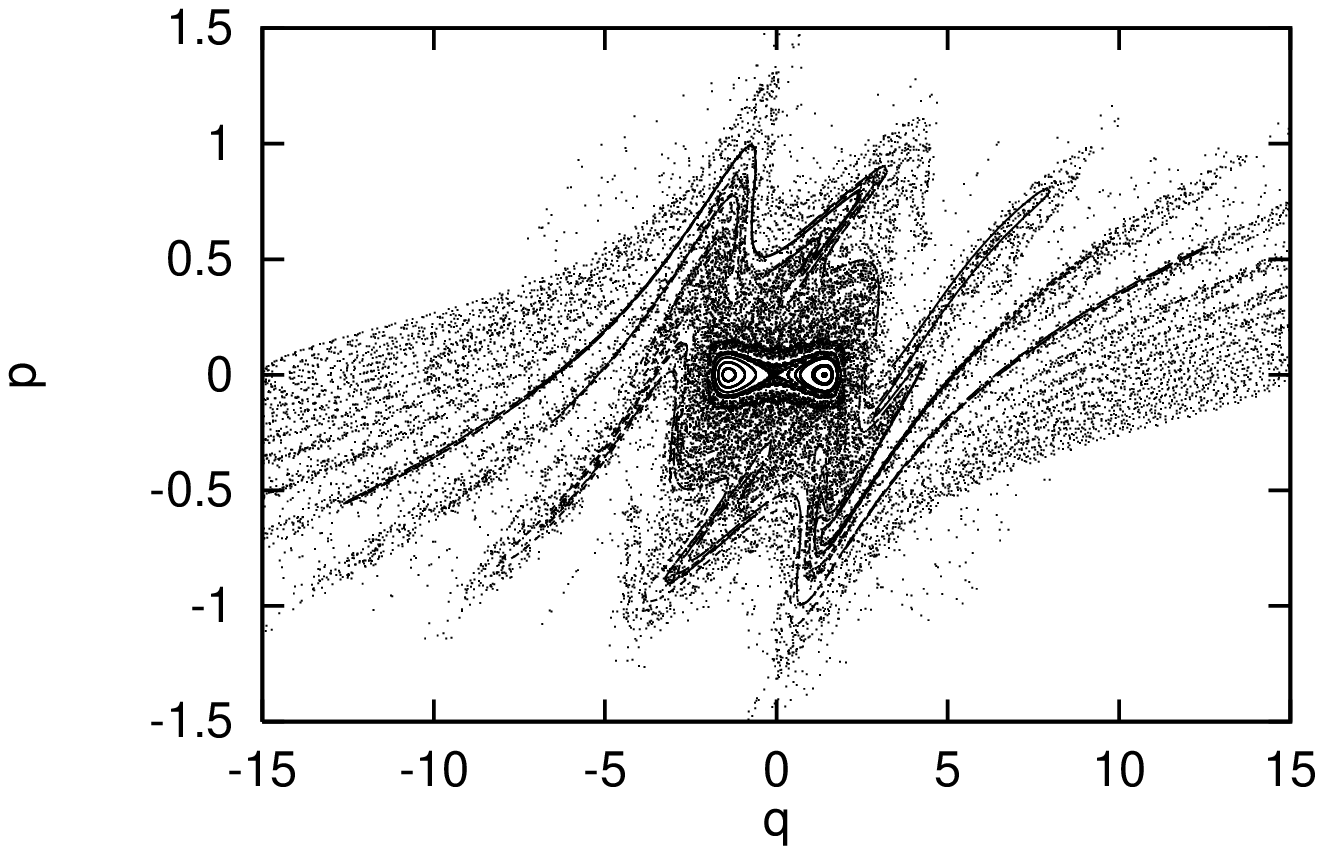}}
\end{center}
\caption{
Poincare surface of section for $ F=5$ and $\omega=1.34$ (Case 2).
Solid line gives the unstable manifolds
}
\label{fig:fig2}
\end{figure}

\begin{figure}[!htb]
\begin{center}
\resizebox *{8cm}{8cm}{\includegraphics*{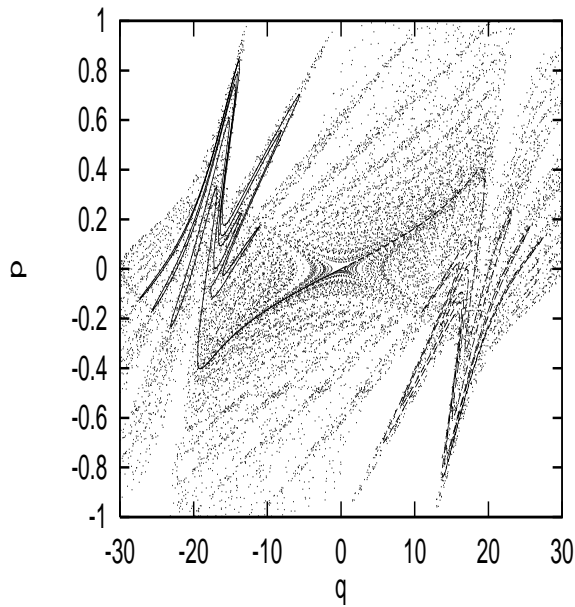}}
\end{center}
\caption{
 Poincare surface of section for $ F=5$ and $\omega=0.52$ (Case 3).
Solid line gives the unstable manifolds.
}
\label{fig:fig3}
\end{figure}

In addition to  those fixed points, KAM tori and Cantori,
another striking structures commonly existing in fig.1,2 and 3
are the
fringe pattern in the lower and upper momentum plane. This is so called
 'fringe structure', first observed in  classical
 simulations of atomic behavior in the stabilization regime by
 Grobe and Law $ [8]$, and recently confirmed by Watson et al in their
 quantum Wigner functions  of quantum description $[6,7]$.

In view of  nonlinear dynamics, the unstable manifold of a steady
state is a set of points $X$ such that the orbit going backward in time
starting from $X$ approaches this steady state. 
The  stable manifold can be similarly defined for the forward orbit. 
A  stable manifold can not intersect other stable manifolds, and 
 neither do the 
unstable manifolds . However,
as chaos occur, stable and unstable manifolds can intersect with each
 other. These invariant manifolds characterize the dynamical properties 
 of the whole system $[9]$. Therefore, 
we conjecture that these fringe structures are relics of the unstable
invariant manifolds. To identify it,
we employ a numerical algorithm (see e.g. $ [10,11]$) to simulate those unstable
manifolds for a rather long time .
 The results are plotted in the same
figure so as to be compared
with the fringe structures. The correspondence between the fringe structures
and unstable manifolds is obvious. 

In case three, the invariant manifolds are obtained according to the
1-periodic hyperbolic point (0,0). However, in case one and case two, things are
much more complicated. Through careful investigations, we find that,
in case one, unstable manifolds of 4-periodic saddles $(P(0.4449, -0.61567),
MP,...M^3P)$ surrounding the stable island associated with the 1-periodic
orbit organize the 'fringe structure'; in case two, the unstable manifolds
of period-6 hyperbolic points $(P(-1.1761,-0.6032), MP,... M^5P)$ around
the fixed point (0,0) construct the fringe pattern.
Because of the symmetry of the map M,
 the stable manifolds and
unstable manifolds are symmetric about x axis.
This fact means that the stable and unstable manifolds intersect
with each other for  infinite  times and form a hyperbolic
invariant set.
Therefore, in the regime of stabilization the classical ionization process
 is characterized by chaos.

\begin{figure}[!htb]
\begin{center}
\resizebox *{8cm}{8cm}{\includegraphics*{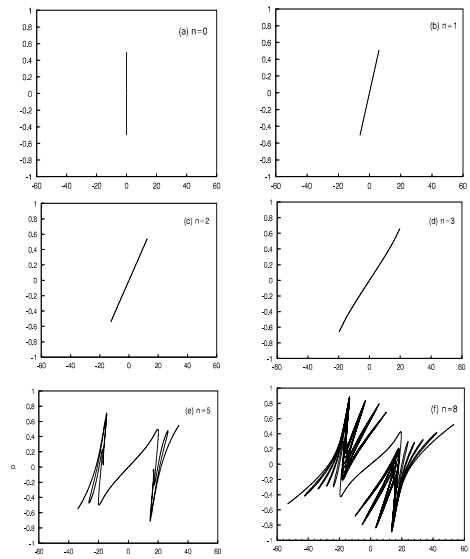}}
\end{center}
\caption{
Phase plots of the $n$-cycle dynamical evolution of
10000 trajectories\\
initiated in a piece-line for $F=5, \omega=0.52$.
}
\label{fig:fig4}
\end{figure}

To deeply understand  the phase dynamics of  ionizaton process in
the classical description,
we trace 10000 trajectories which initiate on a piece-line of the phase plane
described by $x=0$ and  $p \in (-0.5,0.5)$. The phase plane 
representation
of the time-evolution of this piece-line is shown in figure 4.
It is  clear that, 
a classical particle  is   attracted by the  unstable manifolds and comes
close to them along the
stable manifolds, meanwhile   it moves in 
the unstable direction determined by the unstable manifolds. 
Therefore, the whole  piece-line is folded and stretched 
along the unstable manifolds and demonstrates a complex geometric configuration. 
Thus, a particle has more probability to stay near the unstable  
manifolds,
 this  is the
reason for the existence of  fringes and those gaps between them.

From the above analyses, we conclude that, unstable manifolds play an
important role in ionization process  and is the dynamical source
of the striking fringe structure in the classical phase plane.

\section{Quantum Signature }
A stable state commonly refers to a quantal state of atom irradiated
by laser with long lifetime. As the stabilization phenomenon is first
discovered in the case of asymptotically high-frequency field,
the eigenstates of the
time-average Hamiltonian (KH states) are assumed to be  stable
states .

However, in order to expect such phenomenon to persist in the face of various experimental
realities and can be realistically pursued in the laboratory, the
field frequency $\omega $ should be a concrete and finite  rather than
asymptotically high. In this case, as was  pointed out by
Sundaram and Jensen , the high-order fourier coefficients of
the expansion of  time-dependence Halmiltonian is also
important. Therefore, the quasienergy eigenstates of the time
evolution operator $\hat{U}(T)$ will act as the candidates of the stable
states which result in the ionization suppression.

The Floquet operator
is unitary and satisfies following eigenvalue equation,
\begin{equation}
\hat{U}(T)|\Psi_{\lambda}> = e^{-i\lambda}|\Psi_{\lambda}>
\end{equation}
where the eigenphase $\lambda$ is real , $\lambda/T$ is the 
quasienergy (QE). The quasienergy wave function $\Psi_\lambda$ describes
QE state. These QE states
 are obtained by diagonalizing $\hat{U}(T)$ with a large basis of plane
 waves , $|n> = |e^{-ip_n x}>, (n=1,2,...,512)$.

For the operator $\hat{U}(T)$ is time-reversal invariant, let the
 time-reversal operator act on the both sides of
 eigenvalues equation (12), we find that $\Psi_{\lambda}$ and
  $\Psi_{\lambda}^{*}$ are both eigenfunctions with the same
  quasienergy . In general, the bounded states are not degenerate for
  one-dimensional quantal system.
  It is evident that potential $V_0(q)$ is symmetric about zero point.
  Therefore we conclude that
   the bound QE states can be described by real functions and parity is
  a good quantum number.

  In following section, we shall take the second case, where the parameters are
   $F=5, \omega=1.34$, as an example
  for  our discussion. 
 We calculate the lowest two eigenfunctions
 and  find that   the ground state (QE0) possesses  even parity and mainly
 concentrats in the zero point; The 1-excited state (QE1) with odd
 parity  demonstrates dichotomous collection at
 $\pm\alpha_0$.

We use the wave packet propagation method to calculate the evolution 
 of the bound QE states. In this approach, an initial
wave packet is propagated by the Fourier spectral method, which
is directly applicable to the free propagation step of the
Floquet operator, with the impulse delivered once every period.
The free time propagation step is divided into many small intervals
to reduce the error, introduced by boundary conditions.
This kinetic propagation is carried out in the momentum
space, since time evolution reduces to simple multiplication in
that space. The impulsive step is performed in coordinated space
for a similar reason. A Fast Fourier Transform (FFT) routine is used to
transform the wave function between these two spaces.
Since the ionization occurs in our problem, an absorptive filter
in the asymptotic region ($>\pm 50\alpha_0$) is used to avoid the
unphysical reflection of the wave packet from the grid boundary once
it is on its way out. The filter takes the form $[12]$
\begin{equation}
f(x)=1/(1+e^{\beta (x-a)})
\end{equation}
where $a$ is a large distance and $\beta$ is the width of the filter
, the wave function is multiplied by the filter function after each time
step.

The time evolution of ground and 1-excited QE states for the case two
are plotted in figure 5 , they show a good straight line
which indicates an exponential decay $e^{-\gamma t}$. The ionization
rate $\gamma$ of the corresponding QE states is well defined and relates
to the extent of the stability of the QE states.

\begin{figure}[!htb]
\begin{center}
\resizebox *{8cm}{8cm}{\includegraphics*{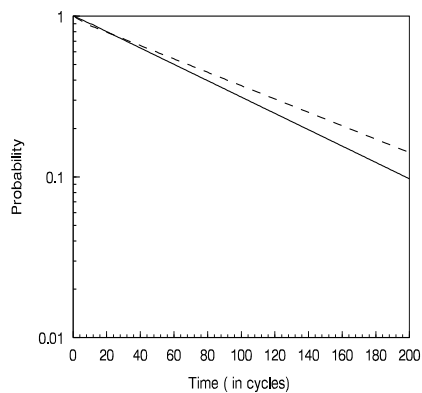}}
\end{center}
\caption{
Survival probability on a logarithmic scale as  a function
of time (in optical cycles)  for $ F=5$ and $\omega=1.34$.
The solid and dotted lines represent the decay of norm of initial
QE0 and QE1 states, respectively, which are calculated from quantum map.
}
\label{fig:fig5}
\end{figure}

To study the phase-space evolution of a wave packet , we use the
 Wigner function which is defined by
\begin{equation}
W(q,p,t) = \int_{-\infty}^{+\infty} \phi(q-\tau/2,t)\phi^{*}(q+\tau/2,t)
e^{ip\tau} d\tau
\end{equation}
for  a given wave function $\phi(q,t)$.
This Wigner function is formally analog to the classical
probability density on phase space, and reduces to it in the limit
$\hbar \rightarrow 0$.

In the exponential decay regime the Wigner function 
is  time invariant
 except for a decay factor. Therefore, we plot the full
Wigner function(a) and its contour plot (b) after 20 cycles of evolution
 in Figs.6,7. The most striking feature of this plot is the 'fringe structure'
 observed in the upper and lower region of the figures. We also use the term
 'fringe' to describe the striking structure observed in the Wigner
 functions
 following Watson et al $[6]$   . 

\begin{figure}[!htb]
\begin{center}
\resizebox *{8cm}{8cm}{\includegraphics*{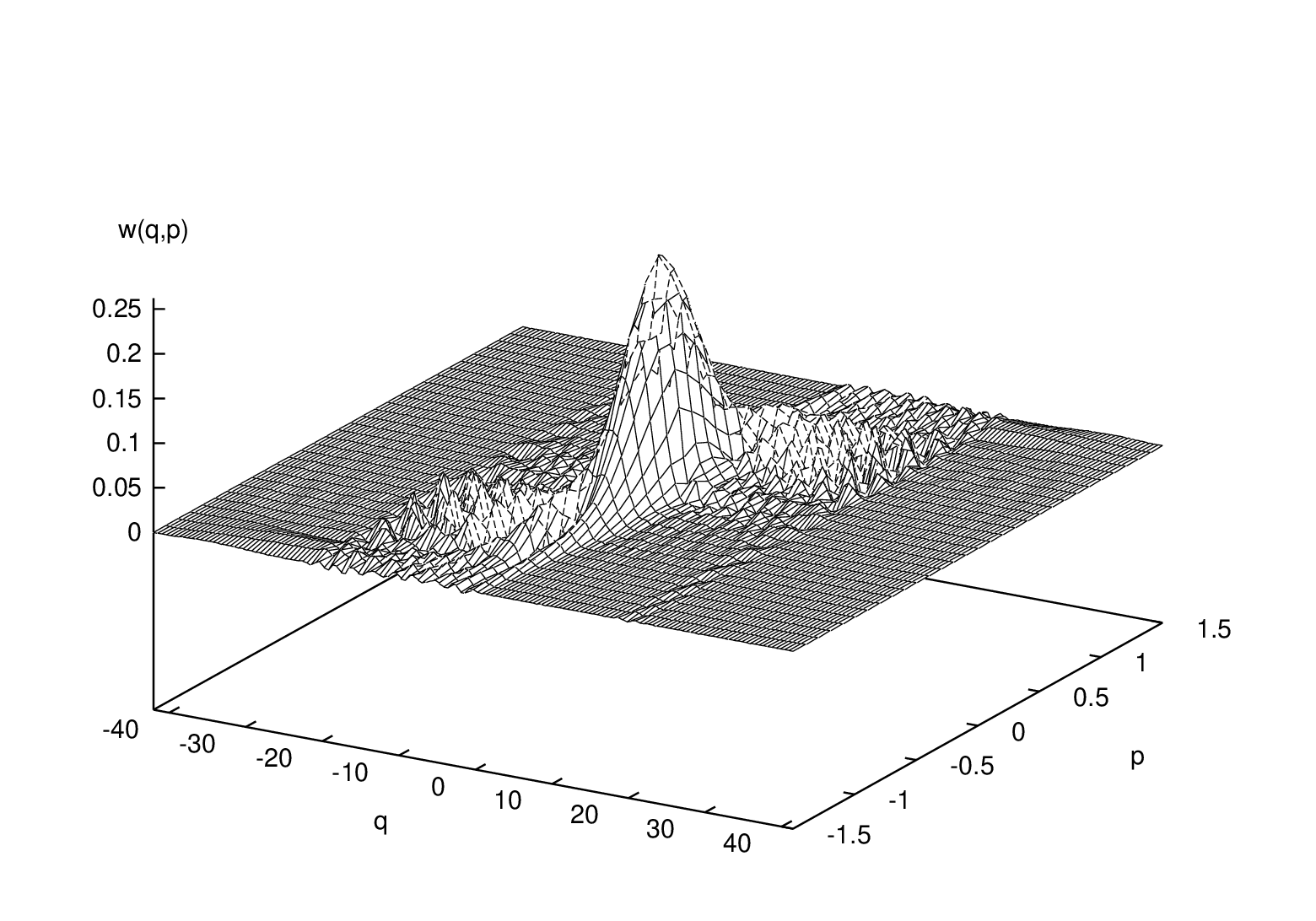}}
\end{center}
\begin{center}
\resizebox *{8cm}{8cm}{\includegraphics*{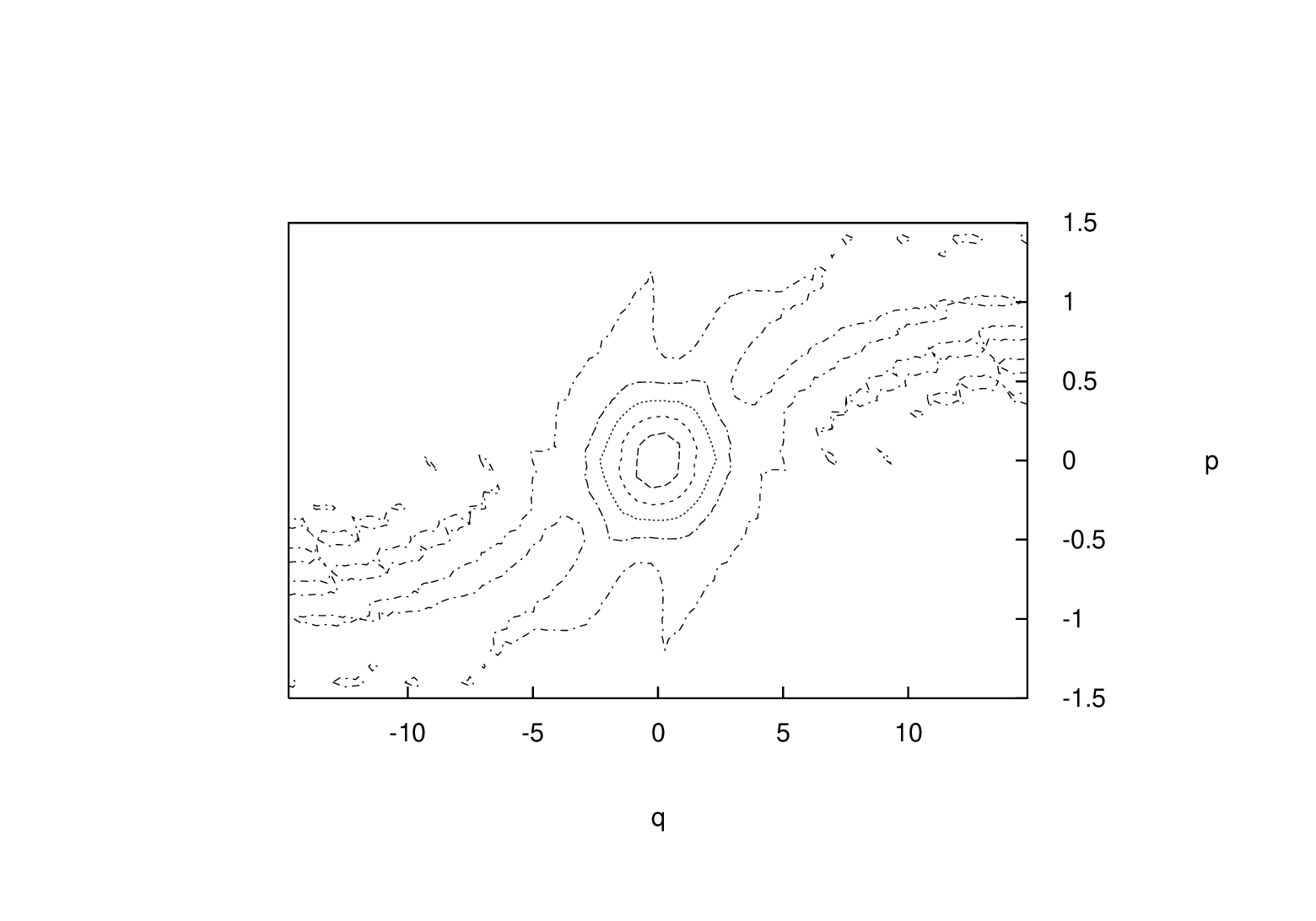}}
\end{center}
\caption{
The Wigner function (a) and its contour plot (b) of
the wave function
after 20 cycles evolution of the QE0 state.
$F=5$ and $\omega=1.34$.
}
\label{fig:fig6}
\end{figure}

\begin{figure}[!htb]
\begin{center}
\resizebox *{8cm}{8cm}{\includegraphics*{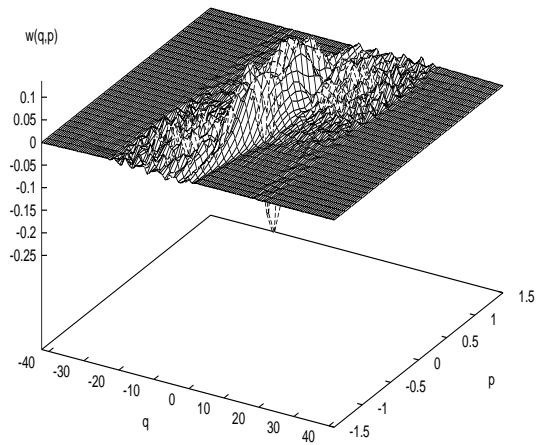}}
\end{center}
\begin{center}
\resizebox *{8cm}{8cm}{\includegraphics*{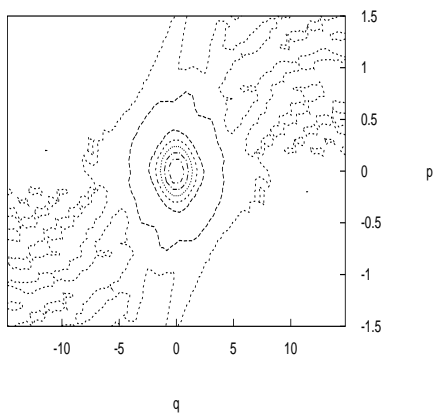}}
\end{center}    
\caption{
The Wigner function (a) and its contour plot (b) of
the wave function
 after 20 cycles evolution of the QE1 state .
$ F=5$ and $\omega=1.34$.
}
\label{fig:fig7}
\end{figure}

\vspace{1cm}
 As to the evolution of the ground QE state, the corresponding Wigner
 distribution contains a strong positive peak at original point and
 many weak positive peaks. The strong positive peak at original point results
 from the even character of the wave function. Those  weak peaks
 resulting in
 the fringe structures
 relates closely with the ionization process.
  In the case of
 1-excited QE state the strong peak at origin is negative because
 of the odd parity of the wave function. In our time-reversal invariant
  model, parity is a good quantum number,
  therefore the Wigner distributions  are
 symmetric about the original point. This symmetry can be observed 
clearly in Figs.6,7.
 When we compare the quantum phase space  with the classical
 unstable manifolds, 
surprisely we find that 
the fringe  structure in classical phase space also shows up.
The fringe structure in both quantum and classical phase space has  the same symmetry,
 and  locates at the same regions.
Since the size of the fringe structures in fig.6 and 7 
is much larger than, the quantum uncertainty, namely,
 the Planck constant $\hbar (=1)$,
 this structure is physically significant. 
So, we believe that it is  
  quantum signature of  the classical unstable manifolds.

  As a further test, we also calculate the Wigner function of the lowest
  two KH states and show them in figures 8,9. Since the wave functions are real and has finite
  parity, they have the reflection symmetry about q axis and p axis.
  Wigner distributions of the ground and 1-excited states contain
  strong positive  peak and strong negative  peak at original point,
  respectively. This character  is the same as that discussed above.
  However, for these bound KH states, no fringe structure is observed.
  This also implies that the fringe structure associates closely with
   the ionization
  process.

\begin{figure}[!htb]
\begin{center}
\resizebox *{8cm}{8cm}{\includegraphics*{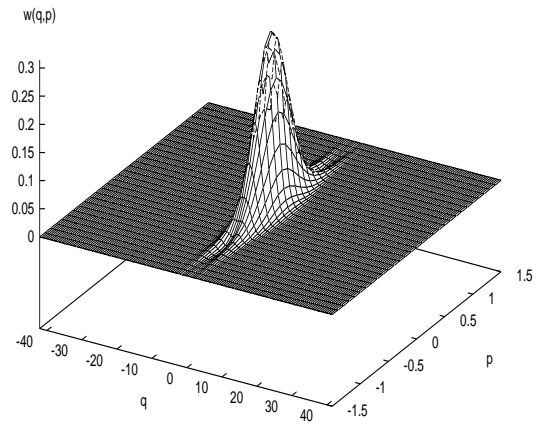}}
\end{center}
\begin{center}
\resizebox *{8cm}{8cm}{\includegraphics*{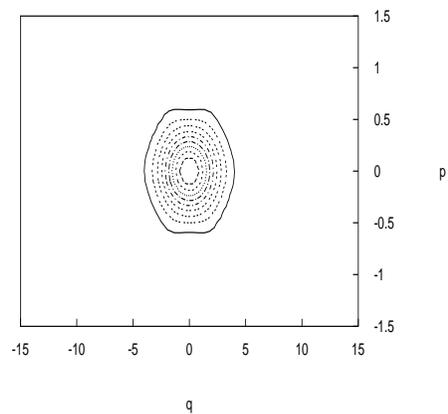}}
\end{center}
\caption{
The Wigner function (a) and its contour plot (b) of
the ground KH state.
$ F=5$ and $\omega=1.34$.
}
\label{fig:fig8}
\end{figure}

\begin{figure}[!htb]
\begin{center}
\resizebox *{8cm}{8cm}{\includegraphics*{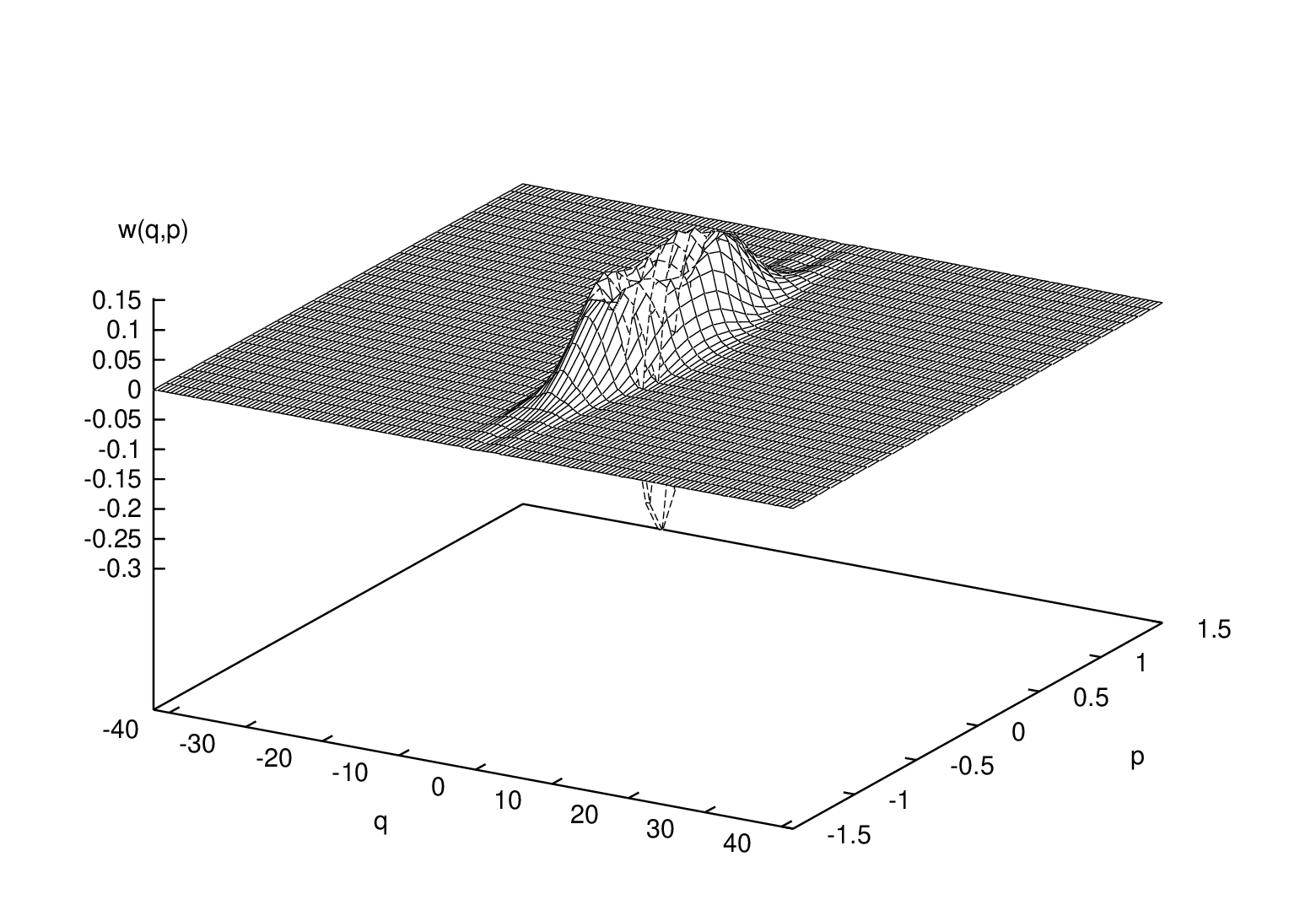}}
\end{center}
\begin{center}
\resizebox *{8cm}{8cm}{\includegraphics*{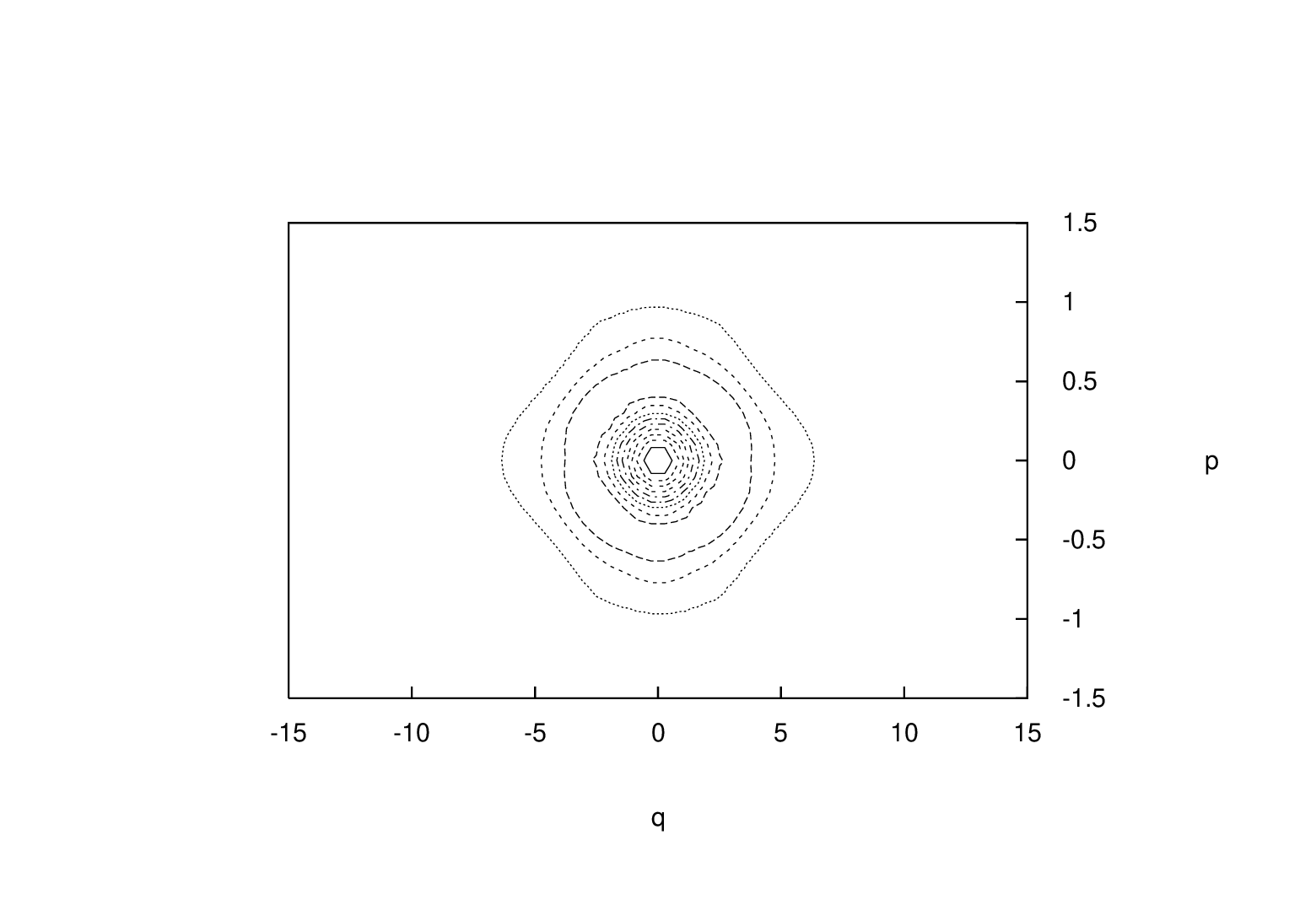}}
\end{center}  
\caption{
The  Wigner function (a) and its contour plot (b) of
the 1-excited KH state.
$ F=5$ and $\omega=1.34$.
}
\label{fig:fig9}
\end{figure}

\section{Conclusions}

In this paper, we have studied the phase-space dynamics of atomic
    stabilization in the frame of a time-reversal invariant model,
  both classically and quantum mechanically. Our results show that  
    1) The dynamical mechanism  of the fringe structure in the
    classical phase space is the unstable manifolds. In this
    situation the ionization process is characterized by chaos.
    2) The fringe structure observed in the quantum Wigner distribution
    is the quantal signature  of the classical unstable manifolds.
    3) The fringe structure has geometric symmetry about
    the original point, and associates closely with the 
 ionization process in the
    regime of stabilization.

Our results show a strong connection between the unstable manifolds and
the ionization process qualitatively, however, it is still an open question 
that how the
ionization rate is related to the fractal structure of the unstable
manifolds, quantatitively. 
 The study along this direction is 
undergoing.

\section*{Acknowledgments}
This work was supported
 by  the Research Grant Council RGC and the Hong Kong
Baptist University Faculty Research Grant FRG
 , and partially
by  the National Natural Science Foundation of China and Climbing Project. 



\end{document}